# L'Equazione del Tempo

**Costantino Sigismondi**
**Università di Roma "La Sapienza" e ICRA, International Center for Relativistic Astrophysics**
sigismondi@icra.it

**Abstract:** the equation of time is fundamental for calculating the rising and setting times of the Sun. The concepts of sidereal and solar day and locak noon, we show how the length of the day depends on solar declination and Earth's orbit position, and also their influence on the local noon shift. In particular we show why in the month of July the setting time is rather constant, while the daylight reduces constantly.

L'equazione del tempo, detta anche analemma è a forma di 8 in posizione orizzontale, è stata usata come simbolo matematico di infinito ∞. La sua conoscenza nei calendari è indispensabile per calcolare gli orari del sorgere e del tramontare del Sole.
Richiamando i concetti di giorno siderale e solare e di mezzodì (§ 1), esaminiamo di seguito come la durata del dì sia influenzata dalla declinazione del Sole e dalla posizione della Terra nella sua orbita (§ 2), e come questi due fattori determinino lo spostamento del mezzodì locale e una variazione non simmetrica degli orari di sorgere e tramontare del Sole (§ 3). In particolare si mostra come nel mese di Luglio l'ora del tramontare del Sole si mantenga pressoché costante, pur riducendosi il tempo di luce.

1. **Il giorno ed il dì.**
1.1 **Giorno siderale.** Definiamo giorno siderale il tempo in capo al quale una stella torna a passare al meridiano locale, cioè a culminare a Sud nella sua orbita quotidiana. Poiché le stelle appaiono fisse nelle loro posizioni sulla sfera celeste, questo periodo corrisponde con il tempo di rotazione della Terra attorno al proprio asse: 23 h 56 m 4 s.
1.2 **Giorno solare.** E' il tempo che il Sole impiega a tornare al meridiano locale. Poiché rispetto alle stelle fisse il Sole si muove verso Est di circa 1° al giorno, in modo da completare il giro delle costellazioni dello zodiaco in un anno, per coprire questo grado angolare la Terra deve ruotare un altro poco, ed il giorno solare è più lungo di quello siderale, e dura in media 24 h.
1.3 **Giorno solare vero.** Poiché l'orbita della Terra è ellittica, la seconda legge di Keplero afferma che attorno all'afelio (4 luglio) viene percorsa più lentamente, quindi il moto angolare del Sole attraverso le stelle fisse è più lento della media, mentre 6 mesi dopo la situazione è opposta. Un Sole più lento corrisponde ad un giorno solare vero più breve (circa 30 s di meno delle 24 h) mentre attorno a 4 gennaio abbiamo il giorno solare vero di ~30 s più lungo. A causa di questa differenza dalle 24 ore nei giorni attorno agli apsidi (afelio e perielio) abbiamo che il Sole torna al meridiano ogni giorno circa 30 s prima (afelio, luglio) o dopo (perielio, gennaio) del giorno precedente. Questa differenza si accumula finché il Sole è più lento (o più veloce) della sua velocità angolare media, fino ad un massimo ritardo di circa 14 minuti a metà Febbraio e un anticipo di 16 minuti ai primi di Novembre. L'anticipo o il ritardo si considera rispetto al mezzogiorno del Sole medio, che per Roma corrisponde con le 12:10 ora solare, poiché si trova a 12°30' di longitudine E da Greenwich, cioè 2°30' W dal meridiano dell'Etna, che corrisponde ad un'ora esatta da Greenwich.
1.4 **Equazione del Tempo.** Sommando all'ora del passaggio al meridiano il valore dell'anticipo (con il segno +) o del ritardo (con il segno -) dato dall'equazione del tempo si ottiene l'ora del mezzogiorno locale medio, e quindi un'indicazione della longitudine. Classicamente la parola *equazione* significava proprio una quantità da aggiungere per rendere una grandezza osservabile uguale ad un' altra da dedurre.

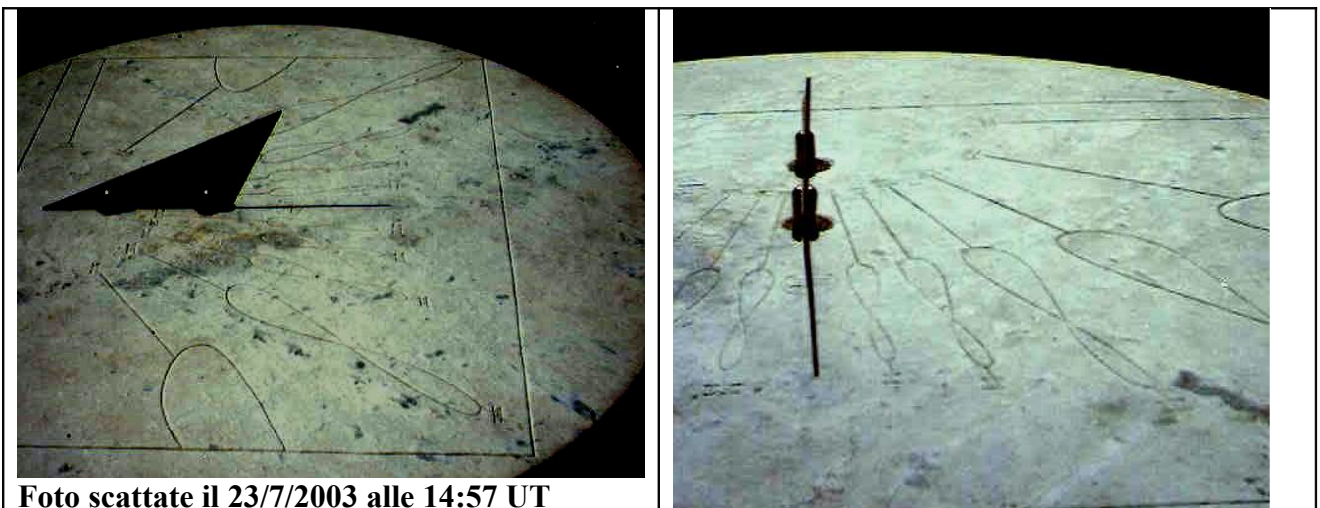

**Foto scattate il 23/7/2003 alle 14:57 UT**

Nella meridiana fotografata sopra abbiamo la rappresentazione degli analemmi per tutte le ore, tranne proprio il mezzodì. Per il 23/7 l'equazione del tempo vale [Duffet-Smith, 1983 p. 77; Barbieri, 1999 p. 49] -6 m 24 s (ritardo). Il mezzogiorno solare medio per quella longitudine cade alle 14:50 UT. Dunque la località si trova circa 42°30' a W di Greenwich. Si tratta, in effetti, di una meridiana a Rio de Janeiro, sulla Praia Vermelha ai piedi del famoso Pao de Azucar. Si noti come l'analemma è rovesciato rispetto a quelli dell'emisfero boreale.

- 1.5 **Il Mezzodì.** E' l'istante rappresentato dalle foto precedenti. Il Sole passa al meridiano locale, ed è trascorso tanto tempo dal sorgere del Sole quanto ne manca al tramonto. Poiché questo istante è soggetto ad anticipare o ritardare a causa della posizione della Terra nella sua orbita, anche l'ora del sorgere e del tramontare del Sole si spostano in conseguenza dell'equazione del tempo.
- 1.6 **Il Dì.** Dal latino *dies*, è il tempo di luce, in cui il Sole è sopra l'orizzonte. La sua durata sarebbe sempre la stessa, 12 ore, se l'orbita della Terra fosse circolare e l'asse terrestre perpendicolare al piano dell'orbita terrestre (eclittica). In termini astronomici ciò accadrebbe se il Sole orbitasse sull'equatore celeste anziché sull'eclittica.

**2. Durata del dì.** Come abbiamo visto il giorno solare vero dipende dalla seconda legge di Keplero; una frazione di questo corrisponde al dì. L'entità di questa frazione dipende dalla geometria, ed in particolare dall'altezza del Sole sull'equatore celeste, cioè dalla sua declinazione.

**2.1 Dipendenza della durata del dì dalla declinazione del Sole.**

La declinazione del Sole la possiamo calcolare in modo approssimato, senza tenere conto della seconda legge di Keplero, considerando l'orbita terrestre circolare. Un moto circolare uniforme è $\theta = 30° \cdot (t - t_0)$, dove t è misurato in mesi a partire dall'equinozio di primavera, $t_0 = 2.69$ che corrisponde al 21 marzo. Sulla sfera celeste un cerchio inclinato di $\varepsilon = 23.5°$, l'angolo di inclinazione dell'asse terrestre sull'eclittica, appare come un'ellisse. L'ellissi si ottiene dal cerchio (di equazioni parametriche $x = 90° \cdot \cos\theta$; $y = 90° \cdot \sin\theta$) mediante una trasformazione del piano che comprime tutte le coordinate lungo la direzione dell'asse minore *a* (la y) del fattore *a/b*, dove *b* è l'asse maggiore dell'ellisse, ovvero il raggio del cerchio di partenza. Nel caso nostro $b = 90°$; $a = 23°.5$. L'equazione della coordinata lungo l'asse minore dell'ellissi ci dà proprio la declinazione solare cercata: $\delta = 23°.5 \cdot \sin\theta$.

Per valutare le ore di luce ad una data latitudine λ, ad un dato tempo t, usiamo la seguente formula (approssimata rispetto al moto Kepleriano, ma giusta sotto il profilo della geometria sferica).

$$T = 12 + \frac{2}{15}\arcsin(\tan\lambda \cdot \tan(23°.5 \cdot \sin(30° \cdot (t - 2.69)))) \qquad \text{[cfr. Sigismondi, 2003 p. 70-75].}$$

La misura del tempo in mesi fa sì che Gennaio va da 0 ad 1; Febbraio da 1 a 2 e così via. In questa formula non si tiene conto della diversa velocità angolare con cui il Sole percorre l'eclittica, ma come se la percorresse a velocità uniforme. Andando più a Nord, d'estate, la durata del dì si allunga, finché ad una certa latitudine il Sole diventa circumpolare (cioè non tramonta mai).

Un' altra formula ancora più approssimata (perché ottenuta in geometria piana) può ottenersi considerando che quando il Sole è sull'equatore celeste percorre in cielo esattamente un cerchio massimo, di cui metà è sotto l'orizzonte (notte) e l'altra metà sopra (dì). Quando il Sole si alza sull'equatore, mezzo cerchio è sempre percorso in 12 ore, anche se è più piccolo del cerchio massimo, ma a questo si aggiunge un pezzo in più all'alba e uno al tramonto. Questi pezzi lunghi ognuno δ·tanλ, vengono percorsi in un tempo complessivo t= 12·δ·tanλ/90° ore. La durata del dì è perciò data da T=12·(1+δ·tanλ/90°)=12·(1+23°.5· sen(30°·(t-2.69)) ·tanλ/90°).

L'ampiezza di variazione della durata del dì con la declinazione solare per la nostra latitudine è di 6 ore all'anno sulle 12 ore di durata media del dì.

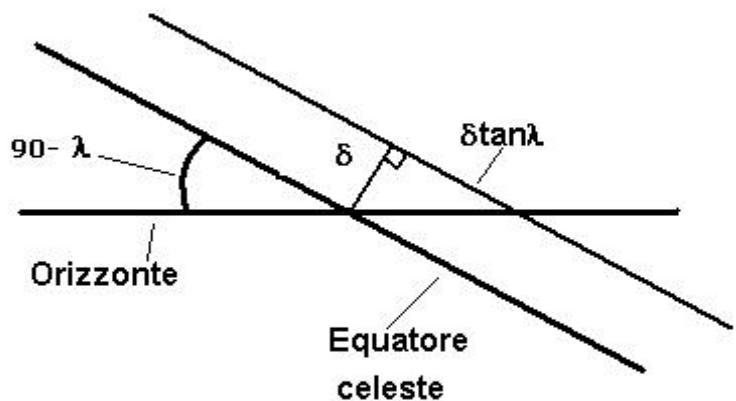

**2.2 Durata del dì a causa della seconda legge di Keplero**
In questo caso la massima ampiezza di variazione è di 52 s su 24 ore del giorno solare medio. Un effetto molto piccolo, dunque, ma che si cumula fino ad oltre un quarto d'ora come abbiamo visto, determinando l'andamento complicato dell'equazione del tempo. Il giorno solare più lungo è a metà dicembre 24 h 00 m 30 s, ed il più corto poco prima dell'equinozio di autunno 23 h 59 m 39 s [Barbieri, 1999 p. 46]. Come si vede tali valori non si presentano, come ci aspetteremmo in prima istanza, esattamente agli apsidi. Questa discrepanza è dovuta al fatto che il Sole, tra un passaggio al meridiano ed il successivo, si sposta anche in declinazione compiendo, specialmente vicino agli equinozi, anche mezzo grado al giorno in declinazione. In realtà la questione della proiezione sull'equatore celeste della distanza angolare percorsa in un giorno sull'eclittica, può essere compresa solo riferendoci alla trigonometria sferica.

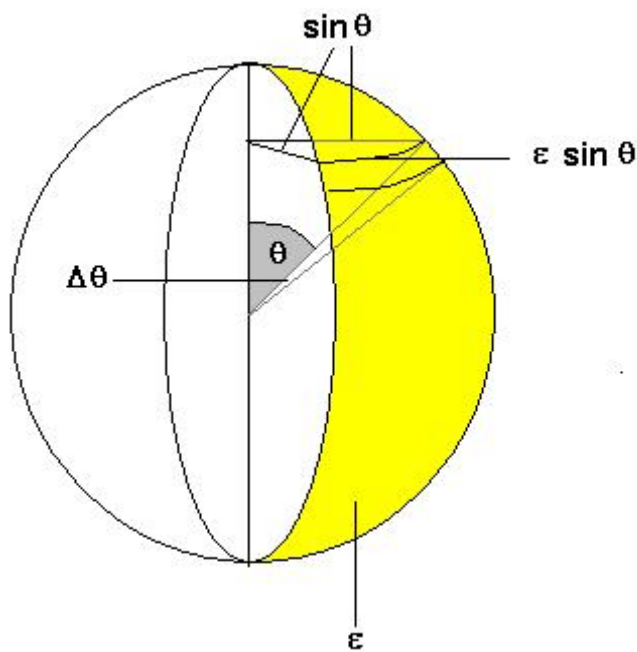

Si vuole sapere l'angolo formato dall'eclittica con l'equatore celeste, per poter calcolare la proiezione su quest'ultimo di ogni arco giornaliero descritto dal Sole vero. Su una sfera prendiamo due cerchi massimi (come i meridiani in figura) e siano essi separati da un angolo ε (espresso in radianti). Un punto che si muova su uno dei due meridiani descrivendo con in centro della sfera un angolo θ ed il suo corrispondente sull'altro meridiano sono separati tra loro di δ=ε·sinθ. Calcolando il rapporto incrementale Δδ/Δθ si ottiene proprio la tangente dell'angolo α tra le due direzioni. α=arctg(ε·cosθ). Quest'angolo si annulla quando i due meridiani tagliano l'equatore, che nella situazione astronomica corrisponde ai solstizi, quando il Sole raggiunge la massima distanza dall'equatore celeste e l'eclittica è parallela all'equatore celeste.

Calcolando questo rapporto dal centro della sfera (come nel caso degli osservatori geocentrici) non occorre calcolare la distanza curva δ=ε·sinθ, ma si usa la retta δ=tgε·sinθ, da cui α=arctg(tgε·cosθ).

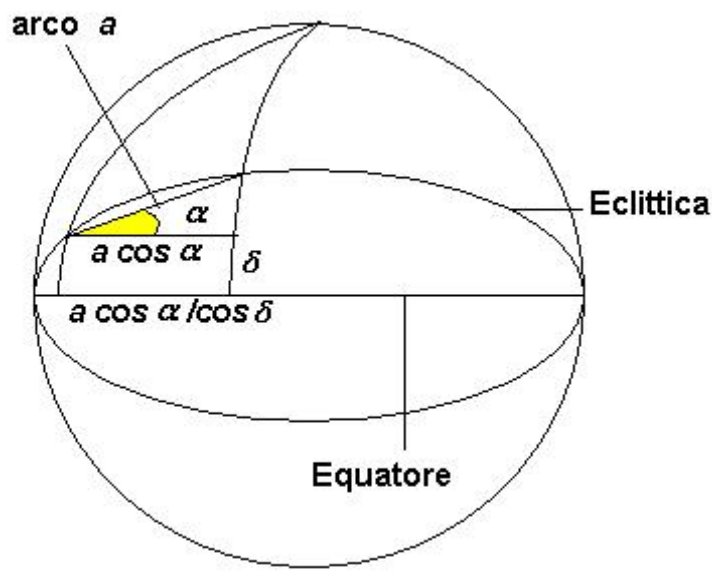

Calcolato α, la proiezione di un arco di eclittica *a* sull'equatore vale *a*cosα/cosδ, poiché la proiezione viene fatta mediante i due meridiani che sono tangenti agli estremi dell'arco *a*, e quindi all'equatore la proiezione si allarga di un fattore 1/cosδ. In questo modo è stata isolata la dipendenza dalla declinazione della proiezione sull'equatore del moto angolare diurno del Sole vero. Per il calcolo del passaggio al meridiano non basta infatti calcolare la proiezione su un cerchio parallelo all'equatore (che sarebbe *a*·cosα nella figura) poiché a declinazioni diverse da zero un astro si muove più lentamente rispetto all'equatore, e la velocità angolare è ridotta proprio di un fattore cosδ. A questo punto abbiamo tutti gli elementi per calcolare la nostra equazione del tempo, ossia i ritardi o gli anticipi del Sole vero rispetto al Sole medio.

**2.3 Sole medio, Sole vero e Sole fittizio.**

La definizione di questi tre strumenti di calcolo astronomico è dovuta a Simon Newcomb, che fu direttore all'Osservatorio Navale degli Stati Uniti a Washington DC a fine '800. Il Sole vero (abbreviato con SV) è quello che orbita sull'eclittica con le variazioni di velocità che gli competono in ossequio alla seconda legge di Keplero. Il Sole fittizio orbita sull'eclittica con una velocità angolare costante, ed è quello a cui ci siamo riferiti per i calcoli sulla durata del dì; il Sole medio orbita, invece, sull'equatore celeste a velocità angolare costante, e passa al meridiano fondamentale di ogni fuso orario tutti i giorni esattamente a mezzogiorno. Il Sole fittizio non passa sempre a mezzogiorno, ma la differenza è calcolabile mediante la proiezione sull'Equatore celeste vista al paragrafo precedente.

La differenza, in gradi, tra Sole fittizio e Sole vero è anch'essa un'equazione, termine inteso come quantità che –da Tolomeo in poi- serve a compensare gli ammanchi di una grandezza calcolata rispetto ad un'osservabile.

Questa equazione è nota come *equazione del centro EC* ed è una soluzione approssimata dell'equazione di Keplero. EC≈115'sin M, dove la cosiddetta anomalia media l'abbiamo espressa in gradi/mese e t è misurato in mesi: M=t·30°/mese. Lo zero di questa M è il 2 gennaio, per i nostri scopi possiamo tranquillamente porlo a t=0.

Che sia una soluzione approssimata è chiaro dalla seguente considerazione: essendo una funzione seno con lo stesso periodo di 12 mesi di quella del Sole fittizio che si vuole correggere, queste correzioni saranno nulle a 3 mesi di distanza dagli apsidi, mentre l'ellitticità dell'orbita implica che il Sole vero ha una velocità angolare minore del Sole medio per poco meno di 6 mesi e maggiore per poco più di 6 mesi, in modo che il raggio vettore Sole-Terra spazzi aree uguali in tempi uguali.

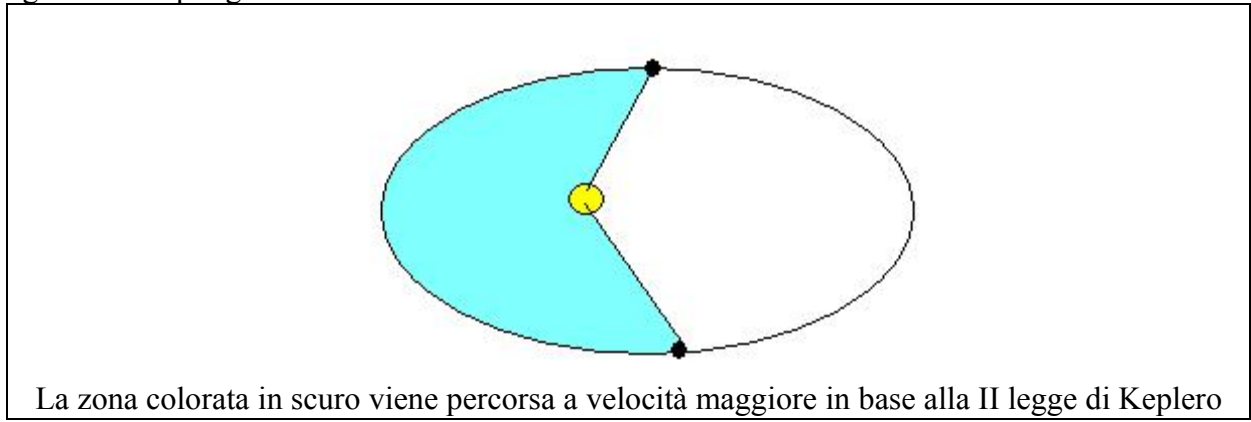

La zona colorata in scuro viene percorsa a velocità maggiore in base alla II legge di Keplero

Incidentalmente questa è la ragione per cui l'estate boreale (afelio) è più lunga dell'inverno (perielio).

### 2.4 Equazione del tempo.

Il Sole fittizio si muove sull'eclittica con velocità angolare costante. Per effetto della proiezione degli archi di eclittica percorsi ogni giorno sull'equatore celeste si ha una differenza variabile tra il giorno solare medio ed il giorno solare determinato da questo Sole fittizio. Tenendo conto anche dell'equazione del centro avremo la differenza tra giorno solare medio e vero.

L'arco medio di eclittica percorso dal Sole vero in un giorno è circa 1°. L'anomalia vera vale $a = [30°·(t-2.69)+EC]$, la declinazione corrispondente è $\delta=23°.5·sen[30°·(t-2.69)+EC]$, la correzione EC va nell'argomento del seno ed è $EC=1°.92·\sin(30°·t)$. La proiezione sull'equatore celeste di questo arco vale infine:

$\Delta\alpha SV=1°·\cos[arctg(tg\varepsilon·\cos(30°·(t-2.69)+EC))]/\cos[23°.5·sen[30°·(t-2.69)+EC]]$.

Questa proiezione è una funzione che oscilla due volte in un anno. L'altra causa di variabilità è la variazione stagionale della velocità angolare: la massima velocità angolare è raggiunta al perielio e la minima all'afelio, una funzione sinusoidale con periodo di un anno ne rappresenta bene la variazione, pur ricordando che si tratta di un'approssimazione in quanto l'orbita è ellittica e valgono le considerazioni del paragrafo precedente.

Possiamo dunque rappresentare la velocità angolare con la seguente funzione adimensionale:

$v=<v>·[1+2\Delta·\cos(30°·t)]$ con $\Delta \approx 0.03 \approx 1.92/30/2$ la semiampiezza della variazione annuale di velocità angolare e $<v>\approx 1$, in modo che il prodotto $v·\Delta\alpha SV$ abbia le dimensioni di un angolo.

L'equazione del tempo ET è data la differenza di $v·\Delta\alpha SV$ dall'ampiezza dell'arco percorso dal Sole medio a velocità uniforme in un giorno. Espressa in frazioni di grado $ET= v·\Delta\alpha SV-1°$, per averla in secondi di tempo basta moltiplicare per 240 (ogni grado angolare corrisponde a 4 minuti di tempo).

Per vedere l'effetto cumulativo del ritardo o anticipo giornaliero si somma algebricamente l'anticipo o ritardo giornaliero ottenendo la curva seguente.

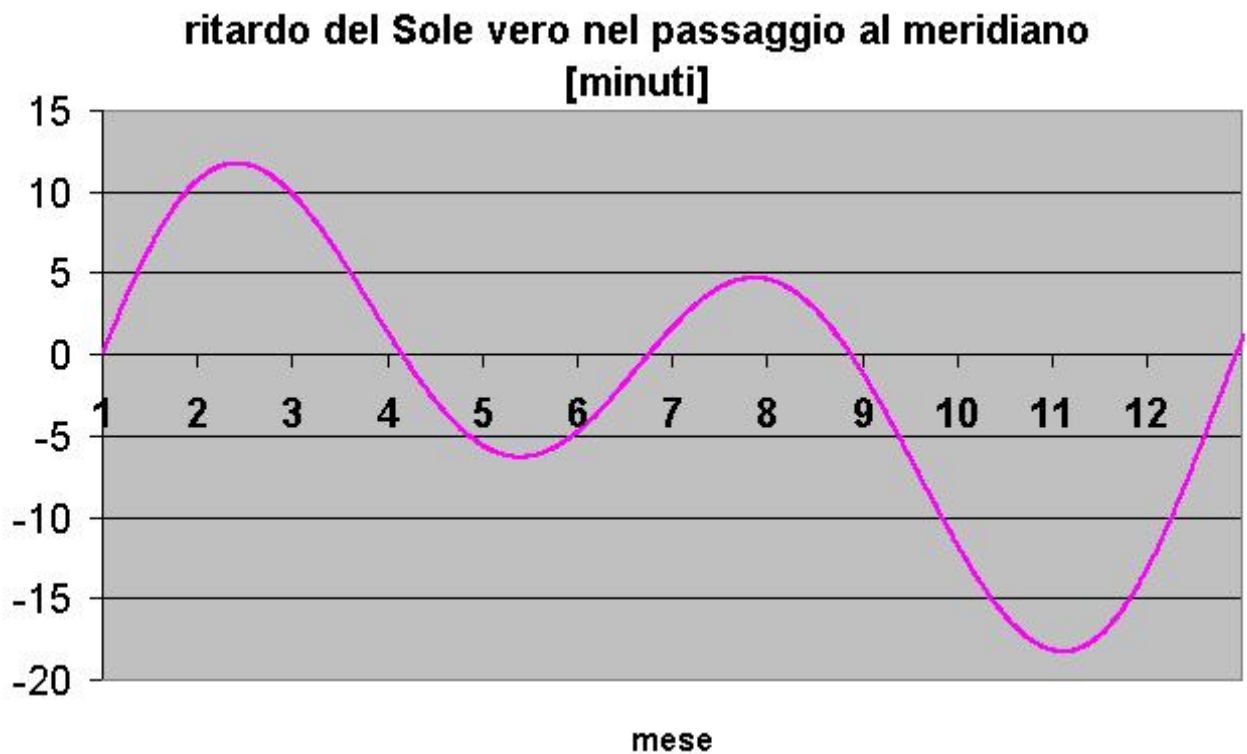

Si noti come in tutte le formule precedenti si sia adottato implicitamente un sistema di riferimento geocentrico, tolemaico. E le formule approssimano tanto meglio l'andamento dei fenomeni, aggiungendo via via termini di armonici superiori ai moti circolari uniformi 'platonici'.

Si è visto anche che nella realizzazione della curva dell'equazione del tempo i coefficienti di questi termini aggiuntivi devono essere noti con una grande precisione, altrimenti la curva assume una forma totalmente diversa e non è più annualmente periodica. Questo significa che nel sistema tolemaico, dove l'equazione del tempo era perfettamente spiegata, i parametri orbitali del Sole (che sono poi quelli della Terra in quello Copernicano) erano già noti con elevata precisione. Si trattava comunque di parametri geometrici relativi e non assoluti, cioè era nota l'eccentricità, ma il raggio dell'orbita non era essenziale alla funzionalità pratica del modello.

### 3. Il Sorgere e Tramontare del Sole

L'equazione del tempo determina lo spostamento periodico del mezzogiorno locale rispetto al Sole medio, il quale passerebbe al meridiano centrale di un dato fuso orario tutti i giorni alle 12 ora solare. Siccome la durata del dì dipende esclusivamente dalla declinazione solare, ne discende che l'ora del sorgere e del tramontare del Sole si trovano sottraendo ed aggiungendo metà dì al mezzogiorno locale o vero, che a sua volta è dato dalle ore 12 con aggiunto il valore dell'equazione del tempo.

Si vede, in particolare, che dall'inizio di Giugno all'inizio di Agosto l'equazione del tempo ha una crescita a velocità costante: il posticipo giornaliero del mezzogiorno vero è costante e vale 1 minuto ogni 6 giorni.

A Roma, 42° Nord, usando le formule del paragrafo 2.1 al principio di Agosto il dì dura 14.3 ore, mentre il 21 Giugno al solstizio estivo dura 15.1 ore. Dal solstizio ad inizio Agosto il dì si riduce di 0.8 ore in 40 giorni, cioè 1.2 minuti al giorno. Questi vanno ripartiti tra alba 0.6 minuti e tramonto 0.6 minuti, ma se il mezzogiorno locale si sposta di 0.17 minuti al dì in avanti, il tramonto del Sole recede alla velocità di 0.43 minuti al giorno, mentre l'alba posticipa di 0.77 minuti al dì.

Così per tutto Luglio il Sole continua a tramontare tardi, anticipando di soli 13 minuti l'ora del tramonto, mentre l'alba ritarda più rapidamente e precisamente di 24 minuti. Nel mese di Luglio sono dunque favorite, in termini di ore di luce, le attività pomeridiane rispetto a quelle mattutine. Questi dati sono ottenuti nell'approssimazione lineare locale sia dell'equazione del tempo che di quella della durata del dì, sufficiente per i nostri scopi.

Si noti che per conoscere la durata esatta della luce del giorno occorre aggiungere ai numeri calcolati circa 3 minuti in più sia all'alba che al tramonto perché la rifrazione atmosferica permette di vedere il Sole un po' prima che tocchi l'orizzonte vero, e un po' dopo che lo ha lasciato.